\documentclass{article}

\begin{document}

\title{On translating conditional statements into mathematical logic}
\markright{translating conditional statements}
\author{Kamaledin Ghiasi-Shirazi}

\maketitle

\begin{abstract}
In this paper, we highlight a profound difference between conditional statements in mathematical logic and natural languages. This difference exists even when the conditional statements are used in mathematical theorems.
\end{abstract}

\noindent
Conditional statements have been the subject of debate in philosophy\cite{anderson1951note, mayo1957conditional, adams1965logic}, mathematical logic \cite{rosen2012, epp2010discrete}, and mathematical education \cite{gregg1997perils, zandieh2014conceptual, inglis2009conditional, durst2019logical}. In this paper, we mention a new subtle point about translating conditional statements to mathematical logic.

How do you translate the English sentence "If both switches A and B are on, then the light L is on" into propositional logic? According to the teachings of logic \cite{enderton2001, rosen2012, johnsonbaugh2018}, you would define the propositions "p: switch A is on",  "q: switch B is on", and "r: light L is on", and translate this sentence as $(p\wedge q)\to r$. 
Now, if I ask you to translate the sentence "either switch A or B or both will turn light L on", then you would write $(p\to r) \vee (q\to r)$.
You probably would be surprised if I tell you that the above two logical expresseions are equivalent and $(p\wedge q)\to r\equiv (p\to r) \vee (q\to r)$ \cite{rosen2012}.
This is surprising as we know that our original English sentences were not equivalent.  

Now, let us promote this paradox to predicate logic. Assume that the predicate $P(x)$ means "x is good", and the proposition $r$ means "the world is paradise".
I ask you to translate the following sentences into mathematical logic:
\begin{itemize}
\item if all humans are good, then the world would be paradise,
\item there exists someone that, the world would be paradise if he is good.
\end{itemize}

According to method taught in logic books, the translations of these sentences to predicate logic are $(\forall x P(x))\to r$, and $\exists x (P(x)\to r)$, respectively.
Again, you would be surprised when you understand that the above sentences are the two parts of the logical equivalence $(\forall x P(x))\to r\equiv \exists x (P(x)\to r)$ \cite{rosen2012}.
This logical equivalence is extremely weird: while the left proposition conditions $r$ on holding $P(x)$ for all objects of the domain of discourse, the right proposition asserts the existence of an object that single-handedly entails $r$.
The reader may think that these paradoxes appear since we are translating everyday English sentences to mathematical logic and won't be a real issue if we confine ourselves to mathematics.
However, these logical equivalences show their specialty even in pure mathematical statements.
It is a common practice to construct proof techniques based on logical equivalences: to prove one side of an equivalence one proves the other side. For example, the equivalence $p\leftrightarrow q \equiv (p\to q) \wedge (q\to p)$ indicates that to prove that two propositions are equivalent, one can prove that each one entails the other. As another example, the equivalence $(p_1\vee ... \vee p_n)\to q\equiv (p_1\to q) \wedge ...\wedge (p_n\to q)$ justifies the technique of proof by cases. If by analogy we try to use the equivalence $(p\wedge q)\to r \equiv (p\to r) \vee (q\to r)$, we end up concluding that to prove that holding both $p$ and $q$ entails $r$, one can equivalently prove that either $p$ or $q$ ,single-handedly, entails $r$; an incorrect consequent.
This shows that, even when stating mathematical theorems, our sense of conditional statements differs from mathematical logic.

The root of these paradoxes is a delicate point about mathematical logic, which is not emphasized in textbooks. A conditional statement in mathematical logic talks about a fixed world in which the truth of the antecedent and consequent are fixed. However, when we use a conditional statement in everyday language, and even when stating a mathematical theorem, what we intend is that the consequent is true in all the worlds in which the antecedent is true. Even in mathematics, we don't formalize theorems whose antecedent is always false.

Assuming that $u$ is a variable that iterates over all worlds (or structure in first order logic terminology \cite{enderton2001}), then our true meaning of the proposition $p\to q$ is that, for every world for which the proposition $p$ is true, the proposition $q$ is true as well, i.e.
\begin{equation}
\forall u [p(u)\to q(u)]
\end{equation}
where we have tacitly assumed that all propositions have an implicit parameter $u$ which designates the situation of the world.
Now, if we come back to the first question, we see that our true meaning of the statement "if both switches A and B are on then the light L is on" is that $\forall u [(p(u)\wedge q(u)) \to r(u)]$.
On the other hand, our meaning of the proposition "at least one of the switches, single-handedly, turns the light on" is that 
\begin{equation}
[\forall u ((p(u)\wedge \to r(u))] \vee 
[\forall u ((q(u)\wedge \to r(u))].
\end{equation}
This new translation clearly shows that the two statements are not equivalent.
If fact, when we talk about the truth of the first conditional statement in mathematical logic, we are talking about a world in which the state of switches and lights is fixed. However, our intention of conditional statements in everyday usage and mathematical theorems is that, in all worlds, the antecedent yields the consequent.
Considering this fact, if we rewrite the statement "if all humans are good, then the world becomes paradise", we have
\begin{equation}
\forall u [(\forall x P(x,u))\to r(u)]
\end{equation}

However, our true intention of the sentence "there exists someone which if he is good then the world is paradise" is $\exists x \forall u [P(x,u)\to r(u)]$. 
The new translations of the English sentences to mathematical logic are no longer equivalent, as desired.
We conclude that in contrast to other logical operators, like $\wedge$ and $\vee$, which usually talk about a fixed situation of a world, in everyday language and even statements of mathematical theorems, a conditional statement means that in all worlds that the antecedent holds, the consequent holds as well.
This is a mathematical interpretation which agrees with our intention of conditional statements in theorems.

\vfill\eject

\end{document}